\documentclass[fleqn,twoside]{article}
\usepackage{espcrc2}

\usepackage{graphicx}
\usepackage[figuresright]{rotating}

\newcommand{\AmS}{{\protect\the\textfont2
  A\kern-.1667em\lower.5ex\hbox{M}\kern-.125emS}}

\title{The interplay of soft and hard contributions\\ in the
electromagnetic pion form factor$^1$}

\author{W. Schweiger\address{Institut f\"ur Theoretische Physik, 
        Karl-Franzens-Universit\"at Graz, A-8010 Graz, Austria}}
       
\begin{document}

\begin{abstract}
We consider various relativistic models for the valence Fock-state
wave function of the pion.  These models are obtained from simple
instant-form wave functions by applying a Melosh rotation to the spin 
part and by imposing physical constraints on the parameters. 
We discuss how the soft and the hard (perturbative) parts of the
electromagnetic form factor are affected by the choice of the model
and by the Melosh rotation.  \vspace{1pc}
\end{abstract}

\maketitle

The knowledge of wave functions of the hadronic constituents allows
to link hadronic phenomena in different kinematical regions.  For
several reasons it is advantageous to consider the composition of
hadrons out of its constituents at fixed light-front time $\tau = t +
z$ rather than at ordinary time $t$.  The ``time'' evolution in $\tau$
is then determined by front-form dynamics.  One of the attractive
features of such an approach is that the corresponding wave functions
are direct generalizations of non-relativistic wave functions in the
sense that they can be interpreted as probability amplitudes for
finding a particular Fock state in the hadron under consideration with
the constituents carrying certain momenta, spins, etc.  Like in the
non-relativistic case light-front wave functions can be expressed in
terms of purely internal variables (momentum fractions $x$ and
transverse momenta $k_{\perp}$). \enlargethispage*{0.7cm}
\fnsymbol{footnote}\setcounter{footnote}{1}
\footnotetext{Talk given at the \lq\lq
International Light-Cone Workshop\rq\rq, Trento, Italy, September
2001}
\begin{table*}[t!]
\caption{Parameters and properties of the wave-function models 
considered.}
\label{table:1}
\newcommand{\m}{\hphantom{$-$}}
\newcommand{\cc}[1]{\multicolumn{1}{c}{#1}}
\renewcommand{\tabcolsep}{0.6pc} 
\renewcommand{\arraystretch}{1.2} 
\begin{tabular}{@{}lrclc}
\hline
wave function           & \cc{$\beta$~(MeV)} & 
\cc{$A_{\pi}$~(GeV$^{-1}$)} & \cc{$P_{\mathrm{q}\bar{\mathrm{q}}}$} & \cc{$\sqrt{\langle 
k^2_{\perp}\rangle}$~(MeV)} \\
\hline
$\psi_{\mathrm{BHL}}$ (full MR)  & \m300 & \m110. & \m1.00 & \m282 \\
$\psi_{\mathrm{BHL}}$ (no MR)     & \m300 & \m78.2 & \m0.51 & \m282 \\
$\psi_{\mathrm{BHL}}$ (approx. MR~\cite{DM87})    & \m300  & \m65.8  & \m1.85  & 
\m323  \\
$\psi_{\mathrm{TK}}$ (full MR) & \m291 & \m77.5 & \m1.00 & \m270 \\
$\psi_{\mathrm{CCP}}$ (full MR) & \m272 & \m80.0 & \m1.00 & \m272 \\
$\psi_{\mathrm{PL}}$ (full MR) & \m1045 & \m238. & \m1.00 & \m266 \\
\hline
\end{tabular}\\[2pt]
$m_{\mathrm{q}}=330$~MeV and $f_{\pi}=93$~MeV throughout.
\vspace{-0.1cm}
\end{table*}

For a hadron form factor the, in principle, exact expression is just a
sum (over all Fock states) of overlap integrals of incoming and
outgoing light-front wave functions~\cite{DY70}.  A widely used
approximation is then to assume that the dominant contribution comes
from the valence Fock state.  For large momentum transfers $Q
\rightarrow \infty$ the analysis of the corresponding overlap integral
reveals that the one-gluon-exchange tail of the wave function can be
factored out, so that one ends up with a perturbative representation
of the form factor in terms of a convolution integral~\cite{LB80}. 
The distribution amplitude $\phi(x, \tilde{Q})$ entering this
convolution integral is again related to the valence-quark light-front
wave function of the hadron.  Its dependence on the factorization
scale $\tilde{Q}$ (which in turn depends on $Q$) is given by an
evolution equation which is driven by one-gluon-exchange. 
Practically, this means that the knowledge of the soft part of the
wave function suffices, since the high-momentum tail of the wave
function is determined by perturbative evolution.  What we therefore
want to model is only the soft part of the pion wave function.

A commonly used ansatz for the quark-antiquark light-front wave function 
of the pion is of harmonic-oscillator type
\begin{equation}
\psi(x,k_{\perp}) = A_{\pi} \chi(x,k_{\perp}) J(x,k_{\perp}) \exp\left(-
\frac{M_{0}^2}{8 \beta^2}\right) 
\label{eq:lcwf}
\end{equation}
with $M_{0}^2$ denoting the front-form expression for the free 
two-particle mass
\begin{equation}
M_{0}^2 = \frac{k_{\perp}^2+m_{\mathrm{q}}^2}{x (1-x)} \, .   
\label{eq:m0}
\end{equation}
$A_{\pi}$ is a normalization constant, $\chi(x,k_{\perp})$ the
(light-front) spin wave function of the q-$\bar{\mathrm{q}}$ pair, and
$J(x,k_{\perp})$ the square root of a Jacobian.  Such a model for the
pion wave function has, e.g., been proposed by Brodsky, Huang and
Lepage (BHL)~\cite{BHL83}.  They took $J=1$ and the usual
(instant-form) expression for the spin-wave function~$\chi$. 
Terent'ev and Karmanov (TK)~\cite{TK76}, on the other hand, chose
$J(x) = \sqrt{1/ 2 x (1-x)}$, i.e. the square root of the Jacobian
relating the relativistic integration measures $(\mathrm{d}^3 k/k^0)$
and $(\mathrm{d}^2 k_{\perp} \mathrm{d}x)$.  The $J(x,k_{\perp})$
adopted by Chung, Coester and Polyzou (CCP)~\cite{CCP88} was of the
form $J(x,k_{\perp})=\sqrt{M_{0}/4 m_{\mathrm{q}} x (1-x)}$, i.e. the
square root of the Jacobian relating the non-relativistic integration
measure $(\mathrm{d}^3 k/m_{\mathrm{q}})$ to $(\mathrm{d}^2 k_{\perp}
\mathrm{d} x)$.  In addition to these different choices of the $J$s we
will use another simple ansatz for the s-state orbital function of the
pion, which has a power-law form
$$
\psi_{\mathrm{PL}}(x,k_{\perp}) = A_{\pi} \chi(x,k_{\perp}) 
J(x,k_{\perp}) \left(\frac{\beta^2}{M_{0}^2+\beta^2}\right)^{\alpha} \, .   
$$

A few words about the spin wave function $\chi(x,k_{\perp})$ are also
in order.  It is a well known disadvantage of the light-front formalism
that the usual addition of angular momenta (e.g. for 2 particles
$\vec{J} = \vec{L}+\vec{S}^{(1)}+\vec{S}^{(2)}$) holds only for the
third component, whereas the addition law for the other two components
is, in general, much more complicated.  It is, however, possible to go
over to a unitarily transformed spin operator which satisfies the
usual spin-addition laws~\cite{TK76}.  This unitary transformation is
known as the ``Melosh rotation'' (MR). The light-front spin wave
function $\chi(x,k_{\perp})$ results thus from an inverse Melosh 
rotation of an ordinary spin wave function. A convenient expression for 
$\chi(x,k_{\perp})$ can, e.g., be found in Ref.~\cite{Ma93}.
Remarkably, the light-front spin wave function of an s-wave pion contains also 
helicity $\pm 1$ components.

The idea behind the construction of such wave-function models is
always that the valence-Fock-state wave function within QCD has
something to do with the wave function of constituent quarks (which
can be obtained from much simpler dynamics).  Therefore the mass
$m_{\mathrm{q}}$ occurring in the wave functions has to be interpreted
as constituent-quark mass.  We took $m_{\mathrm{q}}=330$~MeV.
According to the finding in Ref.~\cite{Sch93} we have also fixed the
parameter $\alpha$ in $\psi_{\mathrm{PL}}$ to be $\alpha=3.5$.  The
normalization $A_{\pi}$ has then been adjusted such that the weak
decay constant $f_{\pi} = 2 \sqrt{3} \int \mathrm{d}x\,
\mathrm{d}^2k_{\perp} \psi = 93$~MeV is reproduced.  Finally, the
parameter $\beta$ has been determined (for the full Melosh rotated
wave functions) by means of the valence-quark dominance assumption
$P_{\mathrm{q}\bar{\mathrm{q}}} = \int \mathrm{d}x\,
\mathrm{d}^2k_{\perp} \psi^{\ast} \psi = 1$.  This also lead to
reasonable values for the mean intrinsic transverse momenta
$\sqrt{\langle k^2_{\perp}\rangle}$ of a quark (or an antiquark)
inside the pion (cf.  Table~\ref{table:1} where the model parameters
are also summarized).

\begin{figure}[t]
\includegraphics[width=7.0cm, clip=]{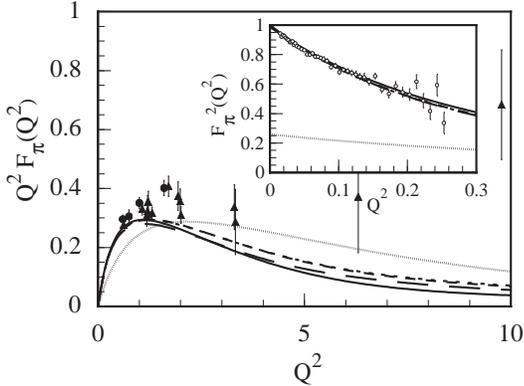}
\vspace{-1.0cm} \caption{Soft contributions to $Q^2 F_{\pi}$ (large
figure) and to $F_{\pi}^2$ (small figure) in different $Q^2$ ranges
for $\psi_{BHL}$ (solid line), $\psi_{TK}$ (short-dashed line),
$\psi_{CCP}$ (medium-dashed line), $\psi_{PL}$ (long-dashed line) with
the full MR factor and for $\psi_{BHL}$ without MR factor (dotted
line).  Data are taken from Ref.~\cite{Be78} (filled triangles),
Ref.~\cite{Vo01} (filled circles), and Ref.~\cite{Am86} (open
circles).} \vspace{-0.7cm}\label{fig:pisoft}
\end{figure}
\begin{figure}[t]
\includegraphics[width=7.0cm, clip=]{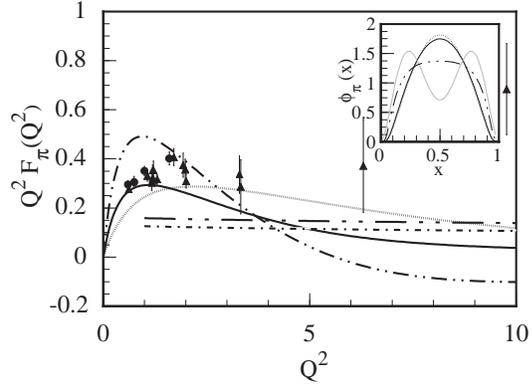}
\vspace{-1.0cm} \caption{Soft (solid line) and hard (dash-dotted line)
contributions to $Q^2 F_{\pi}$ for $\psi_{BHL}$ with the full MR
factor.  Results without (dotted line) and with approximate
treatment~\cite{DM87} (dash double-dotted lines) of the MR factor are
shown for comparison.  Corresponding pion distribution amplitudes
are displayed in the small figure along with a recently proposed
model~\cite{BMS01} (grey line).} 
\vspace{-0.2cm}
\label{fig:pihard}
\end{figure}

The dependence of the soft (overlap) contribution to $F_{\pi}$ on the
choice of the quark-antiquark wave function is displayed in
Fig.~\ref{fig:pisoft}.  At very small $Q^2$ (small figure) the curves
nearly coincide.  Figure~\ref{fig:pihard}, on the other hand, shows
the interplay of soft and hard (perturbative) contributions to
$F_{\pi}$ and the influence of the MR factor.  Results are only considered
for $\psi_{\mathrm{BHL}}$, since the hard contribution to the form
factor is nearly independent of the choice of the wave function.  It
does not even depend on whether the MR factor is fully taken into
account or neglected at all.  Marked differences are only observed if
the MR factor is approximated like in Ref.~\cite{DM87}.  Shortly
summarized, the following conclusions can be drawn from the figures
(and from Table~\ref{table:1}):

    \noindent The use of different wave functions does not have a significant
    effect on the electromagnetic pion form factor as long as these
    wave functions provide comparable results for $f_{\pi}$ and
    $\langle k_{\perp}^2\rangle$.  This holds in both, the soft and
    the hard regimes.

    \noindent Normalizing the wave functions in such a way that
    $f_{\pi}$ remains unaltered, the soft part of $F_{\pi}$ (and also
    $P_{\mathrm{q}\bar{\mathrm{q}}}$) becomes considerably smaller if
    the MR factor is neglected.  Minor changes due to the omission of
    the MR factor can be observed in the perturbative part of
    $F_{\pi}$.  The approximations for the MR factor adopted in
    Ref.~\cite{DM87}, on the other hand, entail also sizable changes
    in the hard part of $F_{\pi}$.

    \noindent Comparing the soft with the perturbative
    contribution to $F_{\pi}$ one observes that the latter starts to
    dominate at $Q^2 \approx 5$GeV$^2$.
    
    \noindent The effect of the MR factor on the pion distribution
    amplitude is just a slight broadening as compared to the
    asymptotic distribution amplitude (cf.  Fig.~\ref{fig:pihard}). 
    The broadening becomes much stronger if the approximations of
    Ref.~\cite{DM87} are applied to the MR factor.  A recent QCD sum-rule analysis of
    the pion distribution amplitude by Bakulev et al.~\cite{BMS01}
    provides a double-humped distribution amplitude (cf. 
    Fig.~\ref{fig:pihard}), but with less pronounced structure than
    originally proposed by Chernyak and Zhitnitsky.

\end{document}